\documentclass[10pt,aps,pre,amsmath,amsfonts,amssymb,floatfix,longbibliography,superscriptaddress,notitlepage]{revtex4-1}
\usepackage{mathrsfs} \usepackage{graphicx,color} \usepackage{bbm} \usepackage{multirow}

\let\a=\alpha  \let\g=\gamma \let\d=\delta
   \let\k=\kappa
\let\l=\lambda    
  \let\f=\varphi 
 \let\y=\upsilon  
\let\D=\Delta \let\Th=\Theta  
   
 \let\r=\rho \let\th=\theta \let\io=\infty

\def\PP{{\cal P}} 
\def\FF{{\cal F}}

 \def\SS{{\cal S}}

  \def\erf{\text{erf}}

\def\de{\mathrm d}

\def\to{\rightarrow}

\newcommand{\beq}{\begin{equation}} \newcommand{\eeq}{\end{equation}}
\newcommand{\wh}{\widehat}

\begin{document}

\title{
Fractal free energy landscapes in structural glasses
} 

\author{Patrick Charbonneau}
\affiliation{Department of Chemistry, Duke University, Durham,
North Carolina 27708, USA}
\affiliation{Department of Physics, Duke University, Durham,
North Carolina 27708, USA}
\affiliation{LPT,
\'Ecole Normale Sup\'erieure, UMR 8549 CNRS, 24 Rue Lhomond, 75005 Paris, France}

\author{Jorge Kurchan}
\affiliation{
LPS,
\'Ecole Normale Sup\'erieure, UMR 8550 CNRS, 24 Rue Lhomond, 75005 Paris, France
}

\author{Giorgio Parisi}
\affiliation{Dipartimento di Fisica,
Sapienza Universit\`a di Roma,
P.le A. Moro 2, I-00185 Roma, Italy}
\affiliation{
INFN, Sezione di Roma I, IPFC -- CNR,
P.le A. Moro 2, I-00185 Roma, Italy
}

\author{Pierfrancesco Urbani}
\affiliation{LPTMS, Universit\'e Paris-Sud 11,
UMR 8626 CNRS, B\^at. 100, 91405 Orsay Cedex, France}
\affiliation{
IPhT, CEA/DSM-CNRS/URA 2306, CEA Saclay, F-91191 Gif-sur-Yvette Cedex, France 
}

\author{Francesco Zamponi}
\affiliation{LPT,
\'Ecole Normale Sup\'erieure, UMR 8549 CNRS, 24 Rue Lhomond, 75005 Paris, France}

\begin{abstract}
\textbf{
Glasses are amorphous solids whose constituent particles are caged by their neighbors and thus cannot flow. This sluggishness is often ascribed to the free energy landscape containing multiple minima (basins) separated by high barriers. 
Here we show, using theory and numerical simulation, that the landscape is much rougher than is classically assumed. Deep in the glass, it undergoes a ``roughness transition'' to fractal basins. This brings about isostaticity at jamming and marginality of glassy states near jamming. 
Critical exponents for the basin width, the weak force distribution, and the spatial spread of quasi-contacts 
at jamming can be analytically determined.
Their value is found to be compatible with numerical observations.
This advance therefore incorporates the jamming transition of granular materials into the framework of glass theory. Because temperature and pressure control which features of the landscape are experienced, glass mechanics and transport 
are expected to reflect the features of the topology we discuss here. Hitherto mysterious 
properties of low-temperature glasses could be explained by this approach.
}
\end{abstract}

\maketitle
Understanding the dynamics of glasses is one of the oldest and most challenging problems in the theory of matter.
The classical landscape picture interprets the slow relaxation of glasses
 in terms of the structures of the free energy landscape: each minimum is a stable amorphous glass state,
high frequency relaxations correspond to vibrational excitations of the state, and slow relaxations
correspond to jumps between different states~\cite{Go69,DS01,Ca09}. 

\begin{figure}
\center{
\includegraphics[width=\columnwidth]{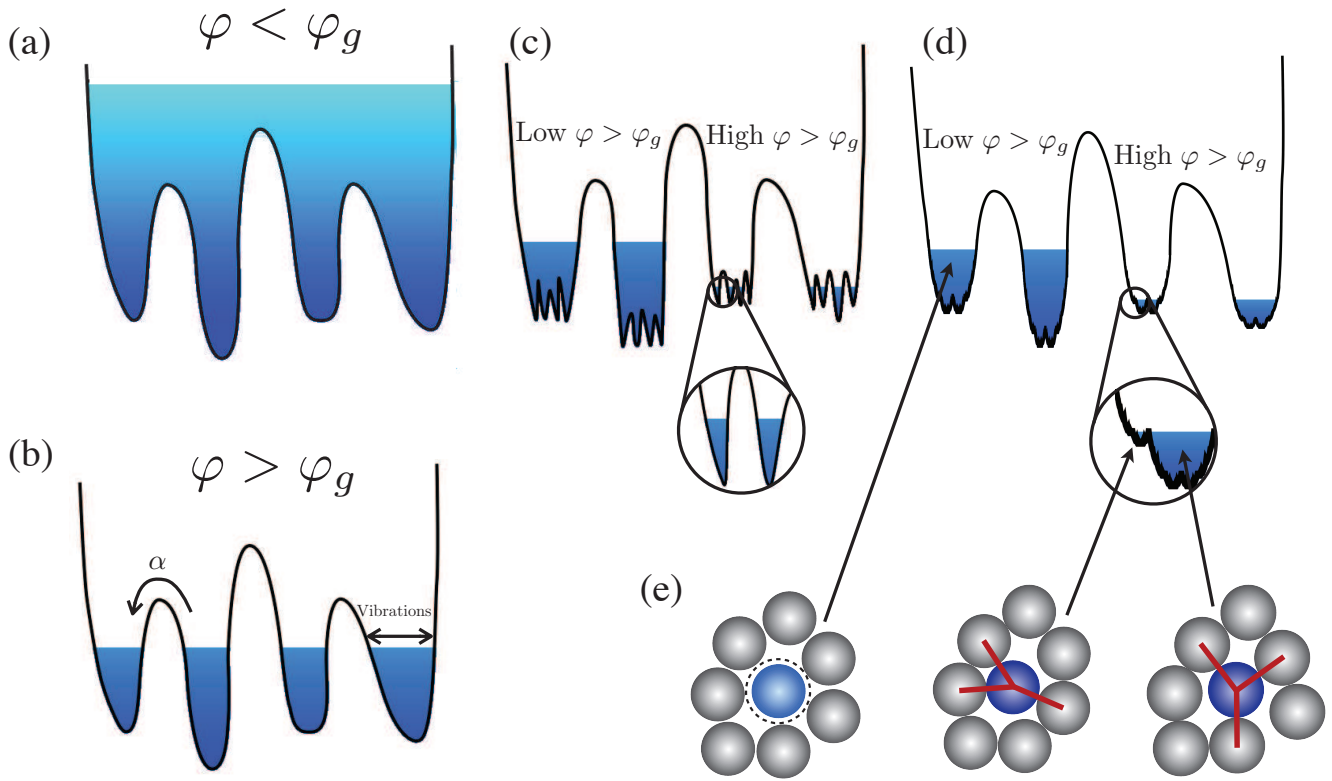}}
\caption{Schematic depictions of (a) the liquid state at packing fractions $\varphi$ that are smaller than the glass transition $\varphi_g$, and of free energy basins for different landscape scenarios: (b) classical stable basins, (c) metabasins of subbasins, (d) and 
metabasins of marginal basins. 
The classical description is akin to boating on a system of lakes separated by high mountains. In the liquid, all of space can be explored. At lower water levels, each basin is a different glass. The free energy barriers hinders passing from one glass to another (the so-called $\a$-relaxation); the basin width allows for vibrational relaxation. 
Both in (c) and (d), the water level further determines what features of the landscape are experienced. Deep into the glass, the landscape roughness results in intra-state barriers that are associated with secondary relaxations.
In (d), at very low water levels (right) -- deep into the fractal glass -- lakes transform into a complex wetland with a hierarchy of small ponds. (e) The very bottom of each of these ponds corresponds to a given realization of the force network (red lines), but the identify of the force contacts remains undetermined before the fractal regime is reached (dashed line).}
\label{fig:basins}
\end{figure}

Yet experimental and numerical observations suggest that this simple landscape description
 -- with essentially only one type of barrier -- is insufficient to capture the complexity of glassy dynamics.
Low-temperature glasses exhibit an intermediate
slow (Johari-Goldstein) relaxation whose timescale is indeed difficult to interpret as corresponding 
to jumps between widely different states~\cite{Go10}. 
It has thus been proposed that the landscape features
narrow subbasins, separated by small barriers, that aggregate into wider metabasins, 
separated by large barriers (Fig.~\ref{fig:basins}). 
Johari-Goldstein relaxation processes would then connect subbasins within a same
metabasin~\cite{HB08,Go10}. 
Direct numerical investigations have confirmed the metabasin organization 
and thereby improved the phenomenological description of transport~\cite{DW08,He08}. 
Deep within the glass phase, the out-of-equilibrium dynamics is also unable to properly sample 
the distribution of barriers associated with the complex subbasin structure, which could explain
why describing it with a single fictive temperature is not possible~\cite{Leheny,Cuku2}.

A different line of evidence for landscape complexity comes from analyzing {\em jammed solids}~\cite{OLLN02,He10,LNSW10}, 
which are amorphous assemblies of hard spheres in mechanical equilibrium. 
Such systems are {\em marginally stable}~\cite{WNW05,WSNW05,BW07,BW09b}: they have very soft vibrational modes and 
excitations that extend over a wide range of timescales~\cite{OLLN02,BW09b}. 
Neither the marginality of the basins, nor the smallness of the barriers associated with the soft modes, fit in the simple landscape picture.

In the late eighties, Kirkpatrick, Thirumalai and Wolynes~\cite{KW87b,KT87,WL12} proposed that mean-field disordered models 
contain the essential features of  glassy landscapes. 
These models come in two broad universality classes: the so-called 
Random First Order (the simple picture of a stable glass, with featureless basins and large barriers)~\cite{KW87b,KT87}, 
and another class where one large state is broken up in a fractal hierarchy
of basins within basins,
discovered by one of us in the Sherrington-Kirkpatrick (SK) model~\cite{Pa83,MPV87}.
 The first class yields, close to the glass transition,
two-step dynamical relaxation~\cite{CK93}, 
in the same universality class as mode-coupling theory~\cite{BCKM97,Go09}. 
It was thus taken to represent
(fragile) structural glasses, at least close to the glass transition -- 
hence the name Random First Order Transition (RFOT) associated 
with this proposal.
Gardner, however, introduced a twist to this classification~\cite{Ga85}. 
She found that, when continued deep in the glass phase, RFOT systems generically reach another phase transition.
At this transition,
each individual amorphous state (basin) becomes a metabasin by breaking 
 into a full fractal hierarchy of subbasins akin to that of the SK model, 
yet retains its identity as a metabasin. 
Surprisingly, despite an early comment to the effect that 
this ``fractal phase'' might be related to secondary relaxations in real glasses~\cite{KW87b}, 
it has since remained somewhat of an intellectual curiosity.

Although the RFOT scenario was initially proposed as 
an analogy, today we know it to be exact    
for particles in the limit of large spatial dimensions $d$~\cite{KW87,KPZ12,KPUZ13}. 
Solving a problem through an expansion around the limit $d \rightarrow \infty$ is an established strategy in
 quantum mechanics, atomic physics and statistical mechanics when there are no small parameters~\cite{1N,GY91}, and the glass problem is no exception. 
The question whether a given feature is captured by RFOT then becomes whether that same feature extrapolates continuously 
from $d=3$ to $d\to\infty$ -- a fact that may be checked with numerical simulations.
It is numerically found that the main features of the bottom of the basins, which are related to jamming, are extremely stable with varying dimension~\cite{SDST06,CCPZ12,GLN12}; note by contrast that the behavior of
high barriers, which are connected to the relaxation around the glass transition, remains the object of lively debates~\cite{DS01,Ta11,BB11}.

The main object  of this paper is to report that  the  {\it exact} hard sphere solution  in the limit $d\to\infty$ 
{\it predicts} the existence of a Gardner transition to a fractal phase in the glass regime, and that 
taking this transition into account 
is crucial to understanding the physics of jamming (Fig.~\ref{fig:basins}).
It affects the out-of-equilibrium dynamics deep in the glass phase~\cite{Cuku2,MR03,MR04,Ri13}, incorporating
(at least partially) secondary relaxations, a point which we here only briefly touch upon. Subbasins 
and barriers of a wide variety of sizes also bring along marginality and soft modes, features that were absent in the original RFOT scenario. 
Their inclusion allows us to make contact with and incorporate the features of jamming theory associated with marginality 
and isostaticity~\cite{WNW05,WSNW05,BW07,BW09b}. More specifically, we show that {\it (i)} the marginal and fractal phase deep 
inside hard sphere glasses fully contains the jamming transition;
{\it (ii)} taking into account this, one can make analytic predictions for the critical exponents of the jamming transition
that are fully compatible with observations; and {\it (iii)} one can compute the probability distribution of the forces in jammed
packings, which displays an analog of the Coulomb gap~\cite{MP07}, 
resulting in a power-law scaling of the distribution of small forces~\cite{CCPZ12,LDW13}. 
Because the critical properties of jamming are independent of spatial dimension~\cite{SDST06,CCPZ12,GLN12}, 
the results obtained in $d\to\io$ immediately translate to experimental systems in $d=2,3$
and hence provide a first unambiguous application of the fractal phase in finite dimension.

\begin{figure}
\includegraphics[width=0.5\columnwidth]{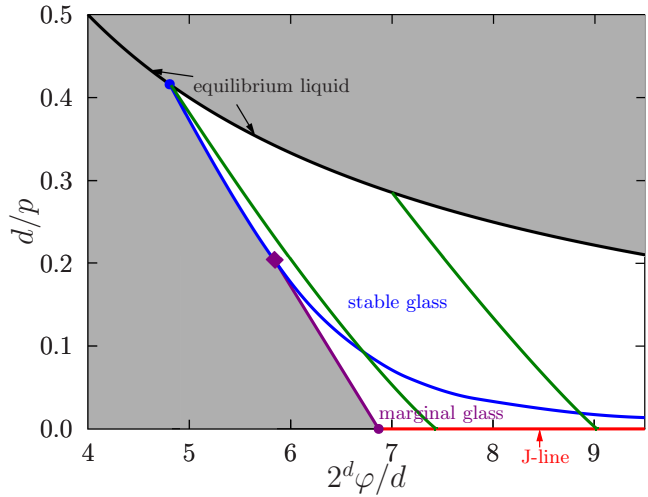}
\caption{Pressure $p$ -- packing fraction $\f$ phase diagram for $d\to\io$ hard spheres. 
The white region indicates the regime where the (meta)basin structure is present, either as a simple stable glass or as a marginal fractal glass. The left-most boundary of the glass region is the threshold line. The ``J-line'' of jammed packings is found along $p=\io$, which always falls within the 
marginal phase. Although solving the mean-field out-of-equilibrium dynamics of hard spheres remains an open problem, an adiabatically slow compression should leave the equilibrium liquid line and eventually reach the J-line, while remaining within the white region. The green lines are two examples of an adiabatic following of a glass state~\cite{PZ10}. 
}
\label{fig:PD}
\end{figure}	

\section*{Results}
\emph{Phase diagram --} 
Using an approach similar to that used for solving the SK model, the exact $d\to\io$ solution for 
$d$-dimensional identical hard spheres of unit diameter
can be formulated in terms of a caging order parameter $\D(y)$~\cite{Pa83,MPV87}. 
This functional order parameter, which encodes the width $\D$ of metabasins on a (properly defined) scale $y$, 
is obtained by numerically solving a set of integro-differential equations (Supplementary Note~1)
and then used to calculate the theoretical liquid-glass phase diagram (Fig.~\ref{fig:PD}). 
The theory predicts that a compressed liquid falls out of equilibrium and becomes a glass at a pressure that
depends on the compression rate. Once in a glass state, further compression results in a quick increase of the system pressure $p$, and upon jamming $p\to \io$~\cite{PZ10}. The final jamming density depends
on compression speed, hence defining a J-line of jammed states~\cite{PZ10}. Two examples of glass compression obtained using an approximate state following are reported in Fig.~\ref{fig:PD}~\cite{KZ10} . 

Independently of compression rate, the glass basin in which the system is initially trapped
undergoes a Gardner transition~\cite{Ga85}, at a line computed in Ref.~\cite{KPUZ13}.
Our key result is that, at pressures above this line, \emph{basins transform
into metabasins that contain a collection of marginally stable glasses}, 
a phenomenon that is described by a non-trivial caging order parameter $\D(y)$ as in the SK model~\cite{Pa83}.
Finding the solution that describes the marginal phase allows us to delimit the marginal phase boundary (Supplementary Note~1) to within the Gardner transition
line of Ref.~\cite{KPUZ13}, the J-line, and the ``threshold'' line determined following the prescription of Ref.~\cite{Ri13}. 
The fact that within this region at least one eigenvalue of the stability matrix in the free energy space vanishes
confirms that this phase is indeed marginally stable (Supplementary Note~1)~\cite{DK83}. We also find that,
while the radius of the innermost fractal basins shrinks to zero as a power-law $\D_{\mathrm{EA}} \sim p^{-\k}$ (see below), 
the radius of the largest metabasins remains of order one.  
Close to jamming the total entropy of a group of metabasins of width $\Delta$ grows as $\Delta^{1/\k}$, 
 hence the basins have a phase space structure whose fractal dimension is $2/\k$ 
(see Supplementary Note~1 for a more detailed discussion).
The {\it marginal} phase is thus also {\it fractal}.

\begin{figure}
\center{
\includegraphics[width=0.5\columnwidth]{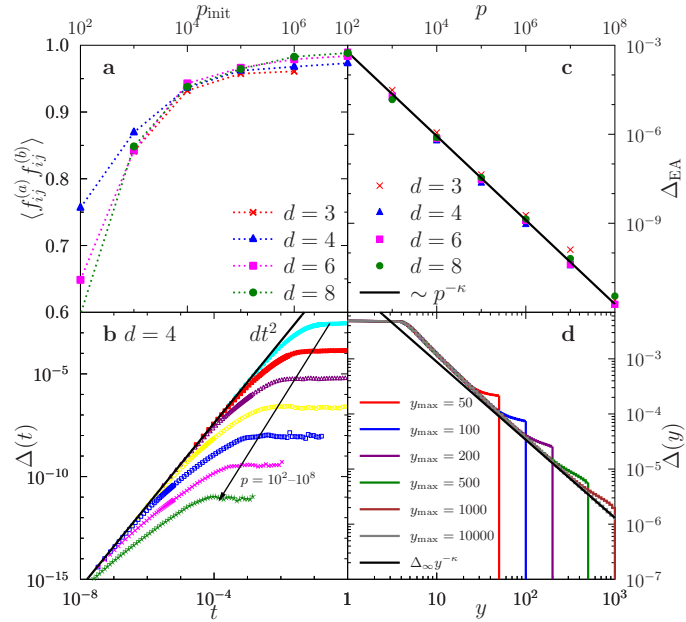}}
\caption{
(a) Overlap between the force network edges $f_{ij}={0,1}$
that connect two particles $i$ and $j$ in two glass configurations $a$ and $b$ (at pressure $p_\mathrm{f} = 10^{10}$),
obtained by independent compression of the same initial configuration at initial pressure $p_\mathrm{init}$ (Supplementary Note~2). 
(b) Time-evolution of the mean-square displacement $\Delta(t)$ for glasses at $p=10^{2}$, $10^3$, $10^4$, $10^5$, $10^6$, $10^7$, and $10^{8}$ in $d$=4. The solid line indicates the ballistic $dt^2$ behavior. The long-time value is the cage size $d \, \Delta_{\mathrm{EA}}$.
(c) The pressure evolution of the cage size $\Delta_{\mathrm{EA}}$ in  various dimension closely follows a power-law $\sim p^{-\k}$ with 
$\k=1.41575$ as predicted by the theory.
(d) Analytical results for the order parameter $\D(y)$ at $\wh\f=10$.
Increasing the cutoff $y_{\rm max}\sim p$ indicates that the scaling regime $\D(y) \sim y^{-\k}$ extends to all
$y$. (Supplementary Note~1)}
\label{fig:MDSA}
\end{figure}

The existence of the marginal phase can be qualitatively tested by molecular dynamics (MD) numerical simulations
in finite $d$ (Supplementary Note~2) by considering the outcome of a slow compression from the liquid up to 
jamming~\cite{CIPZ11,CCPZ12}.
Jammed systems are isostatic, 
and thus particles have an average of 2$d$ force-bearing neighbors~\cite{Mo98,TW99,Ro00,OLLN02,BW09b,CCPZ12}, 
which is much smaller than the $\mathcal{O}(e^d)$ neighbors that isotropically cage a particle in an equilibrated dense liquid. 
Because the identity of the force-bearing neighbors at jamming uniquely characterizes the state,
their emergence sensitively depends on the landscape structure (Fig.~\ref{fig:basins}). 
In the simple basin scenario, force-bearing neighbors at jamming should be fully determined immediately upon leaving the equilibrium liquid; in the meta/subbasin scenario, that determination should only occur once sufficiently deep in the glass for transitions between subbasins to be fully suppressed; in a fractal phase, 
by contrast, the contacts should be gradually determined as jamming is approached. 
To test this scenario, we consider a glass configuration at pressure $p_{\mathrm{init}}$. Starting from this configuration, we perform several
independent compressions up to $p_\mathrm{f} = 10^{10}$ and for each compressed configuration we measure the force network. 
We obtain a set of contact variables $f^{(a)}_{ij}$, which are set to unity if particles $i$ and $j$ 
form a force-bearing contact in configuration $a$ and
to zero otherwise. The average of $\langle f^{(a)}_{ij} f^{(b)}_{ij} \rangle$ over pairs $ab$ of compressed configurations and over contacts $ij$ provides
a measure of similarity between the force networks. The fact that this quantity 
increases smoothly upon increasing $p_\mathrm{init}$ indicates that
the force network is only partially encoded in the initial configuration, in support
of the fractal landscape scenario (Fig.~\ref{fig:MDSA}a and Supplementary Note~2). 

\emph{Criticality of the jamming transition --} 
The equations that describe the marginal phase are formulated in terms of the caging order parameter $\Delta(y)$ and the pair correlation function $g(r)$, which also encodes the probability distribution of forces in the packing. 
Upon approaching the J-line, i.e., as $p\to\io$, these equations develop a scaling regime (Fig.~\ref{fig:MDSA}d) that is characterized by three main critical exponents: $\theta$ for the weak forces, $\alpha$ for the quasi-contacts, and $\kappa$ for $\D$ itself.
A (non-trivial) generalization of the approach developed for the SK model~\cite{Pa06} allows us to obtain theoretical values for these exponents (Supplementary Note~1). Interestingly, the condition that fixes their value is precisely equivalent to the marginal stability condition. 
The theory therefore predicts that the criticality of the jamming transition directly follows from its location inside the marginal phase.

A striking signature of marginality is the scaling of the innermost basin width captured by the Edwards--Anderson cage size 
$\D_{\mathrm{EA}} \sim p^{-\kappa}$. Although $\kappa=3/2$ was proposed in earlier studies~\cite{WNW05,WSNW05,BW09b,IBB13},
the theory predicts a slightly smaller $\kappa = 1.41574$ that is in remarkable agreement with our numerical results 
(Fig.~\ref{fig:MDSA}c).
Because single-particle caging by immediate neighbors (a simple Einstein model for glasses), 
would give $\k=2$~\cite{IBB13}, $\k < 2$ implies that fluctuations near jamming are divergently larger than for independent vibrations, in support of
their cooperative nature~\cite{BW09b,IBB13}. 
Note that if one ignores the fractal phase, an explicit computation erroneously gives $\k=1$~\cite{KPUZ13}.
Also, note that the exponent $\k$ controls the fractal dimension of the basins, as discussed above. 

\begin{figure}
\center{
\includegraphics[width=0.5\columnwidth]{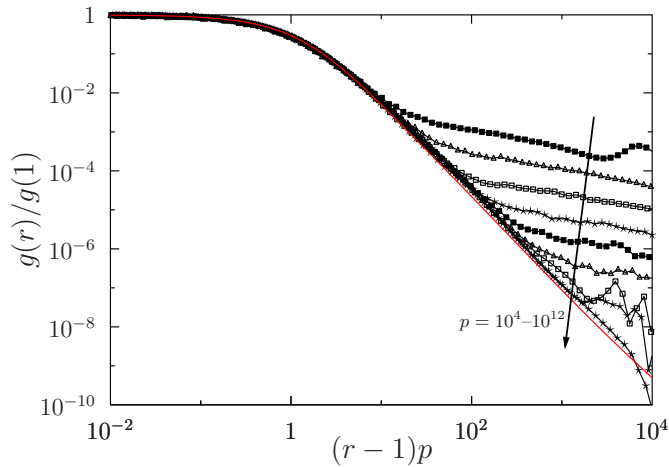}}
\caption{
Scaling of the contact peak of the pair correlation $g(r)$. The theory predicts that in the limit $p\to\io$ one has
$g(r)/g(1) = \FF(\l)$ with $\l = (r-1) p$. Points are numerical data for $g(r)$ in $d=4$ obtained at several pressures 
$p=10^{4},10^5, \cdots, 10^{12}$;
the full line is the theoretical prediction. 
}
\label{fig:gr}
\end{figure}

The pair correlation function $g(r)$ also bears a signature of the criticality at the jamming transition.
The theory predicts, consistently with the analysis of~\cite{DTS05}, that when $p \to \io$, $g(r)$ develops an isostatic contact peak
characterized by a scaling function $\FF(\l) \equiv g(r)/g(1)$ for $\l=(r-1)p$ (Supplementary Note~1 and Fig.~\ref{fig:gr}). 
It also predicts that the scaling
function of the contact peak decays as $\FF(\l)\sim \l^{-2-\th}$ at large $\l$.
The distribution $P(f)$ of inter-particle forces in the packing, which is related to the scaling function of the contact peak by
$\FF(\l)=\int_0^\io df \, f\, P(f) \, e^{-\l f}$~\cite{DTS05,PZ10,CCPZ12}, thus also decays as a power law $P(f)\sim f^\theta$
at small forces. Note that, as observed in~\cite{Wy12}, this phenomenon is closely related to what happens to the distribution of  
frozen fields in the SK model~\cite{SD84}, 
which is thought to explain the Coulomb gap in interacting electron systems~\cite{MP07}.
Beyond the contact peak, the slower decay of pair correlation function follows another power-law $g(r) \sim (r-1)^{-\a}$ that describes the abundance
of quasi-contacts.
These scalings of $g(r)$ are crucial for determining the mechanical stability of packings~\cite{Wy12,LDW13}. Perturbing a packing breaks
some contacts with small forces, while also forming new contacts from what were previously quasi-contacts. Based on this observation a scaling relation for mechanical stability $\a = 1/(2+\th)$ can be derived~\cite{Wy12}. Remarkably, the exponents {\it predicted} by our theory, $\a=0.41269$ and $\th=0.42311$, satisfy this scaling relation to within numerical precision.
Prior estimates of these exponents were also obtained by numerical simulation. The quasi-contact exponent $\a$ has been measured by several groups in dimension $d$ ranging
from $2$ to $13$, all obtaining 
roughly $\a\approx 0.4$~\cite{DTS05,SDST06,CCPZ12,LDW13,AST13}, the most precise estimates being $\a=0.41(3)$~\cite{CCPZ12}.
The weak force exponent $\th$ is, however, more difficult to measure,
and values spanning the interval $\th \approx 0.2 \div 0.45$ have been reported~\cite{CCPZ12,LDW13}. 
Although the existence of a second exponent $\theta'<\theta$ has been shown to affect the tail of $P(f)$~\cite{LDW13}, 
its role in determining $\FF(\l)$ and its large-dimensional scaling remains to be clarified.
Additional numerical simulations are thus needed to test the theory more stringently.

A prediction for the force distribution $P(f)$ at jamming is also available from the theory.
But because this function is not completely determined by the scaling regime, it must be obtained by solving
the full equations that describe the marginal phase. 
Numerically, the function $\FF(\l)=\int_0^\io df \, f\, P(f) \, e^{-\l f}$ is much easier to measure than $P(f)$ because it only depends on structural information, while, in
hard spheres, forces must be determined from the collision dynamics. A measure of $\FF(\l)$ thus also provides a more precise way to measure $\th$.
The theoretical prediction for the scaling function $\FF(\l)$ and therefore for $\th$ are
tested against numerical simulations in Fig.~\ref{fig:gr}, with very good agreement.

\section*{Discussion} 
We have described the marginal phase that is present below
the Gardner transition for hard spheres in $d=\io$.
Using this result we have shown that the jamming transition happens inside the marginal phase and
its criticality in low-dimensional systems
is perfectly described by our approach.
This analysis opens the way for analytically determining many other properties of jammed packings,
such as their shear modulus~\cite{BW06,Yo12}, 
and the properties of avalanches~\cite{LMW10}. 
It also offers a pathway for understanding other glass properties. The nature of aging, for instance, undergoes an abrupt change
at a temperature lower than the glass transition~\cite{St77}, which could signal the crossing between the stable and the fractal glass regimes~\cite{Ri13}.
The observation of dynamical heterogeneities in low temperature glasses~\cite{VKBZ02,BDC08} at timescales much shorter than the inter-basin relaxation 
(a fact that does not fit in the two-timescale picture)  may also be taken as a  signature of
the fractal phase, wherein the dynamical correlation length diverges.
Finally, note that in the $d=\infty$ limit, the barriers controlling the inter-basin relaxation should scale as $d$, 
and the largest separating subbasins as $d^{1/3}$, suggesting the existence of quasi-localized excitations~\cite{XVLN10} 
(stringlike, in three dimensions~\cite{WWW}).

\section*{Methods}

Results are based on the combination of analytical and numerical methods. Analytical results come from the exact solution of hard spheres 
in the limit $d\to\io$, which, for convenience, is obtained 
using the replica method, but any other method would give the same result.
The fractal phase is described by a function $\D(y)$ for $y\in [1,y_{\rm max}]$, as in the SK model~\cite{Pa83}. The cutoff $y_{\rm max} \sim p$ diverges with pressure.
With these definitions $\D_{\rm EA} = \D(y_{\rm max})$ is the mean square displacement in the smallest subbasins, where
$\D(y)$ can be computed by numerically solving a set of coupled integro-differential equations obtained from the replica approach (Supplementary Note~1). Numerical results are obtained by standard event-driven 
molecular dynamics simulations in $d=3$ to 8~\cite{SDST06,CCPZ12}. Compressions are made using
the Lubachevsky-Stillinger algorithm~\cite{SDST06} (Supplementary Note~2)

\emph{Acknowledgments --}
We thank Silvio Franz for an extremely useful discussion. We also thank Ludovic Berthier, Carolina Brito, Florent Krzakala, Edan Lerner, Federico Ricci-Tersenghi, Tommaso Rizzo, Hajime Yoshino and Matthieu Wyart for many interesting exchanges related to this work. Financial support was provided by
the European Research Council through ERC grant agreement
no. 247328 and ERC grant NPRGGLASS. PC acknowledges support from the Alfred P. Sloan Foundation.

\appendix
\section{Supplementary Note 1\\Exact solution of hard spheres in the limit of infinite dimensions}

This supplementary note reviews the exact solution of a hard sphere system in high dimensions, which is obtained through the replica method
in the full replica symmetry breaking (fullRSB) scheme~\cite{Pa83}. We sketch here the main logical steps that allow one to obtain the 
results presented in this work. A detailed derivation will be presented elsewhere.

\subsection{General formulation}

The starting point is the virial expansion for the entropy $\mathcal{S}$ as a function of the density field of hard sphere particles. 
Taking the high-dimensional limit allows us to retain only the first two terms of the virial series~\cite{FP99}
\beq
\mathcal S[\r(x)]=\int \de ^d x\,\r(x)[1-\log \r(x)]+\frac 12\int \de^dx\de^dy\r(x)\r(y)f(x-y),
\eeq
where $f(x)=e^{-v(x)}-1=-\th(D-|x|)$ is the Mayer function and $D$ is the hard sphere diameter. The equilibrium entropy of the system is obtained by solving the stationarity equation $\delta \mathcal S/\delta \r(x)=0$. At low density, the system is in the liquid phase and the solution is $\r(x)=\r$. 
As density is increased the system undergoes a first-order phase transition to a crystalline phase, whose symmetry is only known in low dimensions.
If crystallization is avoided, the system instead enters a metastable supersaturated liquid phase. 
In three dimensions, particular care must be taken both in numerical simulations and in experiments to avoid the crystal phase~\cite{PZ10}, 
but as soon as dimension is increased, crystal nucleation is dynamically suppressed~\cite{SDST06,CIPZ11}, which enables the study and characterization of amorphous states.

The well-established theoretical approach to study these amorphous states consists of coupling the system to a spatial external random field that destabilizes the crystal and favors glassy configurations~\cite{Mo95}. One is then left with the problem of studying $m$ copies (or replicas) of the original system infinitesimally weakly coupled together. 
In order to describe a particle system, this replicated system can be studied by different approximate resummations of the virial expansion, as in standard liquid theory~\cite{MP96,MP09}.
In high dimensions, the virial expansion for the replicated system then reads
\beq\label{replicated_entropy}
\mathcal S[\r(\overline x)]=\int \de \overline x\,\r(\overline x)[1-\log \r(\overline x)]+\frac 12\int \de \overline x\de \overline y\r(\overline x)\r(\overline y)f(\overline x-\overline y),
\eeq
where $\r(\overline x)$ is the single molecule density field \cite{MP09,PZ10} and is a function of $m$ $d$-dimensional coordinates $\overline x=\{x_1\ldots x_m\}$.
The phase diagram can be obtained by changing the parameter $m$ after analytically continuating the expression to non integer values~\cite{Mo95,MP96,MP09,PZ10}. 
The replicated entropy of the system is then determined by solving the saddle point equations $\delta \mathcal S/\delta \r(\overline x)=0$.

Requiring that the solutions of this equation be translationally and rotationally invariant leaves with a function $\r(\overline x)$ that can only depends on the scalar products $q_{ab}=u_a\cdot u_b$, where $x_a=X+u_a$ with $X$ being the center of mass of all $x_a$. A detailed study of this problem provides two important results~\cite{KPZ12}: \emph{(i)} the analytical expression of the replicated entropy in Eq.~(\ref{replicated_entropy}) in terms of the matrix $\hat q$ of the scalar products of the displacement vectors $u_a$ in the infinite dimension limit, \emph{(ii)} the demonstration that the exact solution of the saddle point equation gives the same replicated entropy as that computed within the Gaussian approximation for $\r(\overline x)$.

Let us write down the expression of the replicated entropy that was obtained in this way. We define the Gaussian ansatz for the density field as
\beq
\r(\overline u)=\frac{\r m^{-d}}{(2\pi)^{(m-1)d/2}\det (\hat A^{(m,m)})^{d/2}}\exp\left[-\frac 12 \sum_{a,b}^{1,m-1}\left(\hat A^{m,m}\right)^{-1}_{ab}u_a\cdot u_b\right],
\eeq
where $\hat q= d\hat A$ and $\hat A^{m,m}$ is the matrix obtained from $\hat A$ by eliminating the last column and and the last row. 
The replicated entropy in terms of a rescaled matrix $\hat \a = \frac{d^2}{D^2} \hat A$ is given in Eq.~(45) of Ref.~\cite{KPZ12} as
\beq\label{eq:gauss_r}
s[\hat \a] =\frac{\SS[\hat \a]}N  = 1 - \log\r + d \log m + \frac{(m-1)d}{2} \log(2 \pi e D^2/d^2) + \frac{d}2 \log \det(\hat \a^{m,m}) -  \frac{d}2 \wh \f \,
 \FF\left( 2 \hat \a \right) \ ,
\eeq
where $s$ is the replicated entropy per particle and we have introduced a reduced packing fraction $\wh \f = 2^d \f / d$ that remains finite at the glass transition, even when $d\to\io$.
The function $\FF(\hat v)$ is defined as follows
\beq
\mathcal F(\hat \y)=\lim_{n\to 0}\sum_{n_1,\ldots,n_m: \sum_a n_a=n}\frac{n!}{n_1!\ldots n_m!}\exp\left[-\frac 12 \sum_{a=1}^m \y_{aa}\frac{n_a}{n}+\frac 12 \sum_{a,b}^m \y_{ab}\frac{n_a n_b}{n^2}\right]\:.
\eeq

From the above expression one can study the saddle point equations for the matrix $\hat \a$. By constraining the form of $\hat \a$ one can restrict the parameter space over which to search for a solution. The simplest ansatz is assuming that $\hat \a$ is completely symmetric under replica exchange, i.e.,
$\a_{ab} = \frac12\D (\d_{ab} - 1/m)$, and thus only depends on a single parameter $\D$. It can be shown~\cite{Mo95,MP09,PZ10} that this
structure corresponds to a standard 1-step replica symmetry breaking (1RSB) computation for models with quenched disorder, which is the structure
assumed in the original RFOT scenario~\cite{KW87b,KT87}.
The hard sphere phase diagram obtained in this way was reported in Ref.~\cite{PZ10}. However, 
starting from Eq.~\eqref{eq:gauss_r}, Ref.~\cite{KPUZ13} showed that the 1RSB solution is unstable in some regions of the pressure-density phase diagram. 

\subsection{FullRSB equations}

In the present paper we discuss the results obtained from a fullRSB solution of the saddle point equations, which is expected, by analogy with spin glass
models, to provide the exact solution of the model when the 1RSB solution is unstable. 
The correctness of the fullRSB solution can be proven by studying its (marginal) stability. Here we obtained indications of marginal stability (we studied one of the eigenvalues
and found that it is identically zero in the fullRSB phase)
but we leave a full discussion, i.e., a computation of all eigenvalues, for future work.
In order to illustrate the fullRSB construction we introduce the fundamental object of our theoretical approach, that is the matrix of mean-square displacements defined by
\beq
\D_{ab}=\frac{d}{D^2}\langle (u_a-u_b)^2\rangle=\a_{aa}+\a_{bb}-2\a_{ab}\:.
\eeq
This matrix is analogous to the overlap matrix of the replica solution of mean-field spin glasses \cite{MPV87}. Not only does it encode the order parameter of the system, 
but its structure reflects how the free energy minima are organized~\cite{Pa83,MPSTV84,MPV87}.
The 1RSB ansatz consists of taking a replica symmetric matrix $\D_{ab}=\D$ for $a\neq b$, which corresponds to the replica symmetric form of $\hat \a$ discussed
above. 
The connection with dynamics is as follows. Consider a particle trajectory $x_i(t)$ and define the mean-square displacement
\beq\label{MSDd}
\D(t)=\frac 1N\sum_{i=1}^{N}|x_i(t)-x_i(0)|^2\:.
\eeq
At the 1RSB level we then have
\beq
\lim_{t\to \io}\D(t)=\D,
\eeq
where $t\to \io$ means that we take the limit of the mean-square displacement for times that are large compared to the microscopic timescale but no larger than the lifetime of the metastable state in which the dynamics is trapped.

The fullRSB solution can be constructed as a sequence of $k$RSB solutions in the limit where $k$ diverges. The first step is a 2RSB solution. In this case we divide the values that can be assumed by the replica indices $a$ and $b$ into $m/m_1$ groups each of them containing $m_1$ possible values. We pose that $\D_{ab}=\D_{2}$ if both $a$ and $b$ are in the same group and $\D_{ab}=\D_1$ otherwise. The 2RSB entropy can be obtained by plugging this ansatz in
Eq.~\eqref{eq:gauss_r} and optimizing the result over $\D_1$ and $\D_2$. In order to go beyond the 2RSB ansatz we can construct a 3RSB matrix by dividing each of the $m/m_1$ blocks into $m_1/m_2$ blocks each containing $m_2$ values and by saying that $\D_{ab}=\D_3$ if we are in the same sub block. Iterating this procedure constructs a $k$RSB solution. In the fullRSB limit where $k\to\io$ the matrix $\D_{ab}$ can be parametrized by a continuous function $\D(x)$ over the interval $x\in [m,1]$. Roughly speaking, the ``index'' $x$ (that corresponds to the continuum limit of the indices $m_i$)
selects a given hierarchical level, and this hierarchy describes the hierarchical structure of subbasins sketched in Fig.~1.
The replicated entropy can be written as a function of $\D(x)$ and is given by
\beq\label{entr_full}
\SS_{{\rm fullRSB}}  = 
-  m \int_m^1\frac{\de x}{x^2}\log\left[ \frac{x\D(x)}m+ \int_x^1 \de z \frac{ \D(z) }m \right] 
-   \wh \f \,
e^{-\D(m)/2}\int_{-\infty}^\infty \de h\, e^{h} 
  [1- e^{m f(m,h)}]  \ ,
\eeq
where the function $f$ satisfies the equation
\beq\label{parisi_equation}
  \frac{\partial f(x,h)}{\partial x}=\frac 12\dot\D(x)\left[ \frac{\partial^2 f(x, h)}{\partial h^2}+x\left(\frac{\partial f(x, h)}{\partial h}\right)^2    \right]  \ ,
\eeq
with initial condition $f(1,h)=\log\left(\frac 12+\frac 12\erf\left[\frac{h}{\sqrt{2\D(1)}}\right]\right)$. Equation~\eqref{parisi_equation} was first obtained by one of us
in the solution of the Sherrington-Kirkpatrick model~\cite{Pa83,MPV87} and is the key connection between hard spheres and spin glasses. 

The expressions above give the replicated entropy within a fullRSB ansatz as a function of $\D(x)$. To obtain thermodynamic results, 
we must, however, optimize the function over $\D(x)$. To write (and solve numerically) the stationarity equations, it is convenient 
to introduce in Eqs.~\eqref{entr_full} and \eqref{parisi_equation} a rescaled variable $y=x/m$ and a rescaled function 
$\wh f(y,h)=mf(x,h) = -\frac{h^2 \th(-h)}{2\g(y)} + \wh j(y,h)$, and to introduce
Lagrange multipliers $\wh P(y,h)$ and $\wh P(1/m,h)$ that enforce both the Parisi equation and its initial condition. 
The final variational equations for the fullRSB solution are
\beq\label{eq:Sscaledfinal_cont}
\begin{split}
\D(y) &= \frac{\g(y)}y - \int_y^{1/m} \frac{dz}{z^2} \g(z) \ , \hskip20pt \Leftrightarrow \hskip20pt
\g(y) =  y\D(y) + \int_y^{1/m} \de z  \D(z) \ , \\
\wh j(1/m,h) & = m \log\left[\frac 12\left(1+\erf\left(  \frac{h}{\sqrt{2 m  \g(1/m)}}  \right)\right)\right] + \frac{h^2 \th(-h)}{2 \g(1/m) }  \ ,  \\
  \frac{\partial \wh j(y,h)}{\partial y} &= 
    \frac 12 \frac{\dot\g(y)}y \left[ - \frac{\th(-h)}{\g(y)} +  \frac{\partial^2 \wh j(y, h)}{\partial h^2}
  - 2y   \frac{h \th(-h)}{\g(y)}  \frac{\partial \wh j(y, h)}{\partial h}
  +y   \left(  \frac{\partial \wh j(y, h)}{\partial h}\right)^2    \right] \ , \\
 \wh P(1,h) &= \,e^{-\D(1)/2  -\frac{h^2\th(-h)}{2\g(1)}  + \wh j(1,h) } \ , \\
 \frac{\partial \wh P(y,h)}{\partial y} &=-\frac12 \frac{\dot\g(y)}y e^{-h} \left\{ 
\frac{\partial^2 [e^h \wh P(y, h)]}{\partial h^2} - 2y \frac{\partial}{\partial h} \left[ 
e^h \wh P(y,h)  \left(   - \frac{h \th(-h)}{\g(y)}  + \frac{\partial \wh j(y, h)}{\partial h}\right)
\right]
\right\} \ , \\
\kappa(y) & = \frac{ \wh\f}2  \int_{-\infty}^\infty \de h \, e^h \,  \wh P(y,h) \left(  -\frac{h \th(-h)}{\g(y)}  +   \wh j'(y,h) \right)^2 \ , \\
 \frac{1}{ \g(y) }  &=  y \kappa(y) -
\int_1^{y} dz \kappa(z)
\ .
\end{split}
\eeq
These equations can be solved numerically either by discretizing them on a grid, or
by going to their corresponding finite $k$RSB iterative representation, that is
\beq\label{eq:Sscaledfinal}
\begin{split}
\wh\D_i &= \frac{\wh \g_i}{y_i} + \sum_{j=i+1}^k \left(\frac1{y_j} - \frac1{y_{j-1}} \right) \wh \g_j \ , \\
\wh j(1/m,h) & = m \log \Th\left(  \frac{h}{\sqrt{2 m \wh \g_k}}  \right) + \frac{h^2 \th(-h)}{2\wh\g_k}  \ ,  \\
\wh j(y_i,h) &= \frac1{y_i} \log \left[ \int_{-\io}^\io dz\, K_{\wh\g_i, \wh\g_{i+1}, y_i}(h,z) \,e^{y_i \wh j(y_{i+1},z)}  \right] \ , \hskip20pt i = 1 \cdots k-1 \ , \\
 \wh P(y_1,h) &= \,e^{-\wh\D_1/2  -\frac{h^2\th(-h)}{2\wh\g_1}  + \wh j(y_1,h) } \ , \\
\wh P(y_i,h) &= \int dz \, K_{\wh\g_{i-1}, \wh\g_i, y_{i-1}}(z,h) \,  \wh P(y_{i-1},z) \, e^{z-h} \, e^{ -y_{i-1} \wh j(y_{i-1},z)+y_{i-1} \wh j(y_i,h)} 
\hskip20pt i = 2 , \cdots , k
\ , \\
\wh\kappa_i & = \frac{ \wh\f}2  \int_{-\infty}^\infty \de h \, e^h \,  \wh P(y_i,h) \left(  -\frac{h \th(-h)}{\wh\g_i}  +   \wh j'(y_i,h) \right)^2 \ , \\
 \frac{1}{\wh \g_i}  &=  y_{i-1} \wh\kappa_i -
\sum_{j=1}^{i-1} ( y_j - y_{j-1}) \wh\kappa_j  \ ,
\end{split}
\eeq
where
\beq\begin{split}
K_{\wh\g, \wh\g', y}(h,z) &= \frac{\exp\left[ -\frac{y}2 \left(  \frac{(z-h)^2}{\wh\g -\wh\g'} - \frac{h^2 \th(-h)}{\wh\g} + \frac{z^2 \th(-z)}{\wh\g'}
  \right) \right]}{\sqrt{2\pi(\wh\g -\wh\g')/y}} \:.
\end{split}\eeq
When taking $k$ to be sufficiently large, the results of the discrete $k$RSB equations converge to the continuum fullRSB ones, which is the strategy we employ here.

\subsection{Numerical solution of the fullRSB equations}

It can be shown that the pressure associated with a given glass basin is $p \propto 1/m$~\cite{PZ10,KPUZ13}, and therefore taking the limit $m\to 0$ corresponds
to bringing the system to the jamming limit $p\to\io$. The equations written above
admit a smooth solution in this limit.
To illustrate the fullRSB structure, we numerically solve Eqs.~\eqref{eq:Sscaledfinal} with $m=0$ and a cutoff $y<y_{\rm max}$, as is needed
for numerical purposes. Repeating the procedure for several values of $k$ reveals that the result does not depend on $k$ when $k$ is large (we find $k=100$ to be a good choice).
The result for the numerical solution is given in Fig. 3 of the main text, and it illustrates that $\D(y) \sim y^{-\k}$ when $y$ is large. Because
$\D_{\rm EA} = \D(y_{\rm max})$ and $y_{\rm max} \propto 1/m \propto p$, we conclude that $\D_{\rm EA} \propto p^{-\k}$.

\subsection{Pair correlation function and the Coulomb gap}

It is possible to show that the function $\wh P(y,h)$ is connected to the pair correlation function. In fact, the pair correlation function
$g(r)$ is given, for $r = D(1 + h/d)$ and $d\to\io$, by
\beq
g(h) = \th(h) \int_{-\io}^\io dz \, e^{z-h} \, \g_{\D(1/m)}(h-z) \, \wh P(1/m,z) \, e^{-m f(1/m,z)} \ ,
\eeq
where $\g_\D(x)$ is a centered and normalized Gaussian of width $\D$.
One can show, by a series of scaling arguments, that in the jamming limit and for large $y$, $\wh P(y,h)$ satisfies the following scaling form
\beq\label{eq:Pscal}
\wh P(y,h)  \sim  \begin{cases}
y^{c} p_0( h y^{c} ) & \text{for } h \sim -y^{-c} \\
y^a p_1(h y^b) & \text{for } |h| \sim y^{-b} \\
p_2(h) & \text{for } h \gg y^{-b}.
\end{cases}
\eeq
The first and second regimes are matched by requiring that $p_0(z) \sim |z|^\th$ for small $z$, and $p_1(z\to-\io) \sim |z|^\th$, with $\th =  \frac{c-a}{b -c}$.
Matching the second and third regimes requires that $p_1(z \to\io) \sim z^{-\alpha}$ with $\alpha = a/b$, and $p_2(h) \sim h^{-\alpha}$ for $h\to 0$.

From this scaling, one can show that the exponent $\a$ controls the power-law divergence of the pair correlation function upon approaching 
contact, as numerically observed in several studies~\cite{CCPZ12}.
One can also show that $g(h)$ develops a delta peak upon approaching jamming. The integral of this peak provides the contact numbers
which is equal to $2d$, i.e., jammed packings are {\it predicted} to be isostatic. Upon approaching jamming, the peak is described by a scaling
function~\cite{PZ10}, which is basically the Laplace transform of $p_0(z)$. Hence the exponent $\th$ enters in the scaling of the delta peak, and
is connected to the probability distribution of inter-particle forces $P(f)$~\cite{DTS05,PZ10}. 
In this way one can show that for small $f$, $P(f) \sim f^{\th}$. 

The exponents $a,b,c$ are determined by imposing three consistency conditions on the scaling regime. 
We can further show that $\k=1+c$.
This analysis will be reported elsewhere. Here we only quote the final result which is
\beq\label{eq:exp_num}
\begin{split}
&a=0.29213\ldots \hskip30pt b=0.70787\ldots
\hskip30pt 
c=0.41574\ldots
\ , \\
&\a = 0.41269\ldots \hskip30pt \th = 0.42311\ldots
\hskip30pt 
\k = 1.41574\ldots
\ , 
\end{split}\eeq
where the precision is given by the last digit.

\subsection{Fractal structure of the basins}
The function $\D(y)$ can also be used to look directly at the structure of hierarchically organized states \cite{MPV87}. In the 1RSB picture of the glass transition, at the dynamical or mode-coupling transition point the liquid minimum of the free energy landscape breaks down into an exponential number of minima organized according to their free energy (here, internal entropy). From this landscape one can study the number of metastable states having internal entropy~$s$
\beq
\mathcal N(s)=\mathrm e^{N\Sigma(s)},
\eeq
where $N$ is the size of the system. In the fullRSB picture the state structure is organized in a hierarchical way. 
 Suppose that we are able to sample configurations from a given state $a$. We then introduce the mean-square distance between two of these configurations, labeled $a$ and $b$, as $\D_{ab}=\frac{1}{N}\sum_{i=1}^N |x_i^{(a)}-x_i^{(b)}|^2$. 
To be more concrete, let us ``lump'' all the states that are at a mutual distance less then $\D$ within a metabasin in which the total  
internal entropy is $s$. We can then try to determine what is their number $\mathcal N(s,\D)$. 
In this way we obtain a coarse-grained description of the configurational entropy at the scale $\D$ that is defined by
\beq
\mathcal N(s,\D)=\mathrm e^{N\Sigma(s,\D)}\:.
\eeq
From this definition it follows that $\Sigma(s,\D(1))=\Sigma_{\rm L}(s)$, where $\Sigma_{\rm L}(s)$ is the configurational entropy of the largest metabasins.
The connection between $\Sigma(s,\D)$ and the mean-square displacement profile $\D(y)$ can be obtained as follows.
We first highlight the dependence on $m$ of the profile $\D(y;m)$, and we introduce the inverse function $y(\D;m)$. The standard 1RSB expression that relates the configurational entropy of metabasins having internal entropy $s$ to $\D$ is then given by
\beq\label{level0}
m y(\D(1;m);m)=m=\frac{\partial \Sigma(s,\D(1;m))}{\partial s} = \frac{\partial \Sigma_{\rm L}(s)}{\partial s} \ ,
\eeq
where we have used the fact that $y(\D(1;m);m)=1$.
This relation holds for the most coarse-grained version of the configurational entropy. If we reduce the coarse graining on states we have on a fixed scale $\D$
\beq\label{generic_level}
m y(\D,m)=\frac{\partial \Sigma(s,\D)}{\partial s} \:.
\eeq
Equation~\eqref{level0} relates $m$ to the total  
 internal entropy $s$ of the metabasins we are looking for, on a fixed scale of metabasin width $\D$. 
 
 It can be shown~\cite{MPV87,MPSTV84} that at equilibrium the metabasins' entropies on a scale $\D$ are {\it independent random variables} distributed
 according to $\PP_\D(s) \propto e^{m y(\D;m) s}$, and therefore the typical value of $s$ on a scale $\D$ is $s_{\rm typ}(\D) \propto 1/y(\D)$. From this result
 it follows that close to jamming, there is a large region of (small) $\D$ where $y(\D) \sim \D^{1/\k}$. In this region, when increasing $\D$ from the smallest
 $\D_{\rm EA}$ to the larger values, one finds that the total basin size grows as $s_{\rm typ}(\D) \sim 1/y(\D) \sim \D^{1/\k} \sim \sqrt{\D}^{2/\k}$
 (recalling that $\D$ is the squared distance between configurations). 
 Hence the basins in phase space form a fractal with dimension $2/\k = 1.41267\cdots$.

\subsection{Out of equilibrium dynamics}
 Following a quench  from high temperature or, in the case of hard spheres, from low pressure,  at initial time $t=0$, a macroscopic system relaxes without reaching equilibrium. 
 During this process, the two-time correlation function
 \beq\label{MSDtwotime}
\D(t,t_w)=\frac 1N\sum_{i=1}^{N}|x_i(t)-x_i(t_w)|^2
\eeq
 never becomes a function of $(t-t_w)$. Instead, relaxation becomes progressively slower as $t_w$ becomes larger, i.e. as the system
 {\em ages}.
 Rather surprisingly, one may infer from a static calculation 
 some features of the out of equilibrium dynamics.
 This follows from two different facts: \\
{\em (i)} The constant pressure dynamics relaxes up to a density level at which the free energy landscape disconnects into separate basins.
This {\it threshold} level~\cite{Cuku} is also the one at which the basins lose their stability, and become marginal.   \\
 {\em (ii)} At long times, the almost-stable states visited by the dynamics are sampled with equal   probability. This is
 a property that emerges in the mean-field out of equilibrium solution~\cite{Cuku,Cuku2}, of which there is no clear general explanation yet.\\
If we compute the   function $\D(y)$ with the value of $m$ fixing it not by maximization of the free energy (minimization of entropy) as in equilibrium,
but rather by demanding that  for $m=m_{\rm th}$ the stability of the solution be marginal (the replicon associated with the largest value of $\Delta$ in the Parisi ansatz vanishes), it may be shown that properties {\em i)} and  {\em ii)} imply that the  values of pressure, free energy, etc coincide with 
 the  asymptotic ones of the dynamics.
  The fluctuation-dissipation ratio is then given by $m_{\rm th}y(\D;m_{\rm th})$, when the time-dependent mean-square displacement is $\D(t,t_w)=\D$~\cite{Cuku2}.

\section{Supplementary Note 2\\Molecular dynamics simulations}

Molecular dynamics simulations of $N$=8000 monodisperse hard spheres  in $d$=3, 4, 6, and 8 evolving in a cubic box under periodic boundary conditions are performed using a modified version of the event-driven code described in Refs.~\onlinecite{SDST06,CCPZ12}. Hard spheres of unit diameter $D$ and unit mass $m$ naturally express time $t$ in units of $\sqrt{\beta mD^2}$ at fixed unit inverse temperature $\beta$. Glasses are obtained from low-density fluids using a Lubachevski-Stillinger algorithm with, in $d>3$, a slow particle growth rate of $\dot{\gamma}=3\times 10^{-4}$~\cite{SDST06}. Compacting the fluid makes it fall out of equilibrium near the dynamical transition and brings the resulting glass arbitrarily close to jamming~\cite{CCPZ12}. In $d$=3, a rapid initial growth with $\dot{\gamma}=3\times 10^{-2}$, in order to prevent crystal formation~\cite{CCPZ12}, is followed by a slower growth rate once the system is well within the glass $p\gtrsim 10^{3}$.

Using these glass configurations as starting point for fixed-density simulations, the mean-square displacement $\Delta(t)= \frac{1}{N}\sum_{i}\langle|x_i(t)-x_i(0)|^2\rangle$ is obtained (Fig.~3 from the main text). Rattlers  are removed from the averaging. They are identified by further compressing the glass up to $p=10^{10}$~\cite{CCPZ12} and identifying particles with fewer than $d+1$ force contacts, i.e., pair distances that are smaller than $D+100/p$.
The long-time plateau then gives $\lim_{t\rightarrow \infty}\Delta(t)/d=\Delta_{\mathrm{EA}}$, where the Debye-Waller factor $\Delta_{\mathrm{EA}}$ estimates the average cage size in the glass. 

The force network overlap $\langle f_{ij}^{(a)}f_{ij}^{(b)}\rangle$ is obtained by performing a pairwise comparison between 100 configurations ($a$ and $b$) that have been independently compressed from the same initial glass configuration. At the final pressure, particles $i$ and $j$ that are within a distance $D+100/p$ from each other are considered to be part of the force network and ascribed a unit variable $f_{ij}$. All other particle pairs are given a null value

\bibliography{HS,glass}

\end{document}